\documentclass[superscriptaddress,nofootinbib, amsmath,amssymb,preprintnumbers,prdfloatfix,twocolumn,PRL]{revtex4-2}
\usepackage{CJK}
\usepackage{graphicx}
\usepackage{dcolumn}
\usepackage{upgreek}
\usepackage{amsmath}
\usepackage{bm}
\usepackage[colorlinks=true,pdfstartview=FitV,breaklinks=true]{hyperref}
\usepackage[dvipsnames,table]{xcolor}
\hypersetup{urlcolor=BlueViolet,
	    citecolor=Plum,
	    linkcolor=PineGreen}
\usepackage{lipsum}	    
\usepackage{titlesec}
\usepackage{etoolbox}
\usepackage[normalem]{ulem}

\newcommand{\dndt}{\frac{\mathrm{d}n}{\mathrm{d}t}}
\newcommand{\rhodm}{\rho_{\mathrm{DM}}}
\newcommand{\fMyr}{f_{\mathrm{Myr}}}
\definecolor{offblue}{RGB}{23,80,153}

\begin{document}

\preprint{IPMU23-0016}

\title{Explaining the GeV excess with exploding black holes}

 \author{Zachary S. C. Picker}
 \email{zpicker@physics.ucla.edu}

\affiliation{Department of Physics and Astronomy, University of California Los Angeles,\\ Los Angeles, California, 90095-1547, USA}

\author{Alexander Kusenko}
\email{kusenko@ucla.edu}

\affiliation{Department of Physics and Astronomy, University of California Los Angeles,\\ Los Angeles, California, 90095-1547, USA}
\affiliation{Kavli Institute for the Physics and Mathematics of the Universe (WPI), The University of Tokyo Institutes for Advanced Study, The University of Tokyo, Chiba 277-8583, Japan}

\begin{abstract}
\noindent 
Black holes may form in present-day collapse of microscopic structures of dark matter.  We show that, if microstructure black holes (MSBH) with mass $m\sim 10^{13}~g$ are produced, the spectrum of gamma rays from their evaporation agrees remarkably well with the GeV excess observed by Fermi Gamma-ray Space Telescope, while still avoiding all observational  constraints. We also discuss the generic requirements for MSBHs to explain the GeV excess.
\end{abstract}

\maketitle

\noindent The source of the Galactic Center gamma-ray excess~\cite{Atwood_2009,Goodenough:2009gk,vitale2009} has been in debate for some time, with explanations ranging from annihilating dark matter to millisecond pulsars, to the supermassive black hole in the center of the galaxy, to high energy cosmic ray sources~\cite{Goodenough:2009gk, Petrovic:2014uda, Brandt:2015ula,Carlson:2014cwa,Gaggero:2015nsa,Fermi-LAT:2017opo,Leane:2019xiy, Cholis:2021rpp,Calore:2021jvg}. All of these explanations still have significant theoretical or observational uncertainties, and the true origin of the GeV excess is still unknown~\cite{Cholis:2021rpp}. 

Small black holes, evaporating via Hawking radiation~\cite{Hawking:1974rv,Hawking:1974sw}, can produce high-energy gamma rays. One may ask whether primordial black holes (PBHs)~\cite{pbh,Hawking:1971ei,Carr:1974nx,Chapline:1975ojl}, which form in the early universe, could perhaps then be the  explanation for the GeV excess. Indeed, the tightest constraints on the abundance of PBHs in the mass range $m\sim 10^{15}-10^{17}~g$ come from requiring the Hawking gamma-ray flux to be smaller than observed in the Galactic Center~\cite{DeRocco:2019fjq,korwar_updated_2023,carr_constraints_2021}. Unfortunately, the gamma-ray spectra from black holes with masses $m\gtrsim 8\times10^{14}~g$, which have lifetimes long enough to survive until today, peak at insufficiently high energies to explain the GeV signal~\cite{mosbech_effects_2022}.
However, we now know of late-time  mechanisms to form black holes besides stellar collapse. If black holes form from the collapse of dark matter microstructures or from first-order phase transitions in the dark sector today~\cite{flores_primordial_2021,chakraborty_formation_2022,lu_late-forming_2022,kawana_primordial_2022,Domenech:2023afs,Picker:2023ybp}, it is possible that a small but rapidly evaporating population of black holes could exist in the Galactic Center. The microstructure black holes (MSBHs) can be very small without running into conflict with observational constraints. 

\begin{figure}[!ht]
    \centering
    \includegraphics[width=\columnwidth]{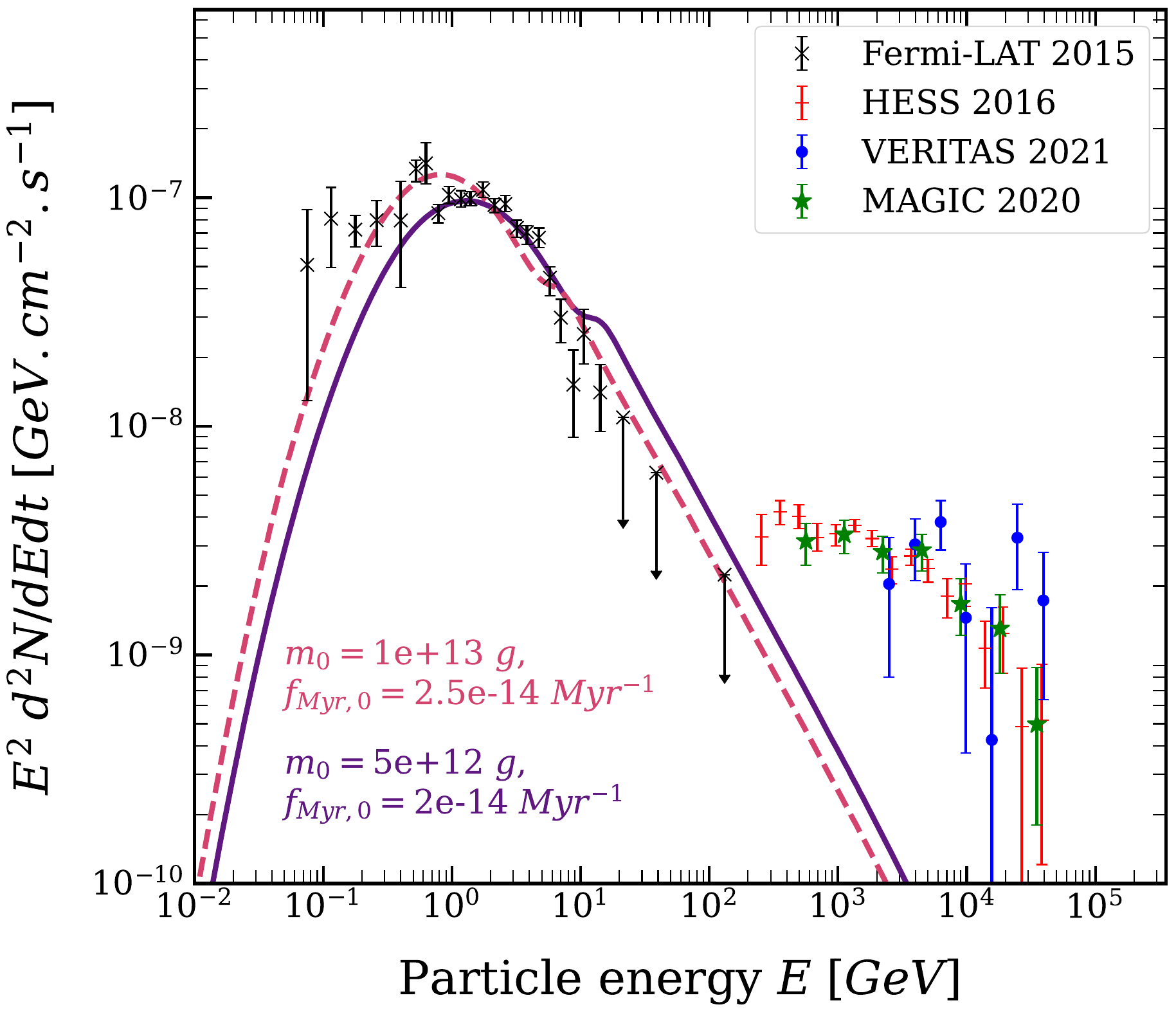}
\caption{Differential energy spectra from exploding black holes in the Galactic Center, calculated at an angular size of one degree radially. The solid purple ($m=5\times10^{12}~g$) and dashed pink ($m=10^{13}~g$) lines give the spectrum for two slightly different choices of parameters for the MSBH model, with $p_m=0$ and $p_f=1$ to account for the signal morphology. Fermi-LAT~\cite{fermilat2015ApJS..218...23A,Malyshev:2015hqa} observations are plotted, with the peak at $\sim1~GeV$ matching the expected signal well. HESS~\cite{hess_collaboration_acceleration_2016}, VERITAS~\cite{Adams:2021kwm}, and MAGIC~\cite{magic2020A&A...642A.190M} observations are plotted based on Fig.~5 from Ref.~\cite{Adams:2021kwm}. The TeV spectra is well fit by the expected gamma-ray spectrum from a Galactic Center PeVatron which is accelerating protons up to $PeV$ energies~\cite{HESS:2017tce,Adams:2021kwm}.}\label{fig:gcenter_spectrum}
\end{figure}

In this {\it letter}, 
we show qualitatively that the resulting gamma-ray spectrum agrees well with observations of the GeV excess by gamma-ray observatories such as Fermi-LAT~\cite{fermilat2015ApJS..218...23A,Malyshev:2015hqa}, as seen in Fig.~\ref{fig:gcenter_spectrum}.  We also outline the generic requirements for a population of MSBHs to explain the GeV excess. We leave a more sophisticated statistical analysis of this scenario for future work, when we have a more complete and detailed model of dark structure formation and evolution to MSBHs---the aim of this short \textit{letter} is merely to demonstrate the possible consistency of exploding black holes with the Galactic Center GeV excess.


There are three basic requirements which a scenario must satisfy in order for black holes to explain the GeV excess: (i) exploding black holes form at late times; (ii) The abundance and size of the black holes are consistent with the observed spectrum; (iii) The distribution of MSBHs is consistent with the observed morphology of the GeV signal. 


The first requirement (i) is related to the second requirement (ii) because, in order to correctly match the GeV excess, very light black holes need to be evaporating. Primordial black holes (PBH) cannot satisfy these requirements without violating stringent constraints on their abundance~\cite{carr_new_2010,Acharya:2020jbv,carr_constraints_2021,green_primordial_2020,DeRocco:2019fjq,korwar_updated_2023,mosbech_effects_2022}. In addition, there would need to be significant fine-tuning in the original population in order to reproduce today's signal---since the evaporation rate is exponential, the mass today is sensitive to very small perturbations in the initial mass, even for extended mass functions~\cite{mosbech_effects_2022}. As a result, we require that a population of exploding black holes be continually injected into the galaxy, i.e. from the late-time collapse of dark matter microstructure. 

We follow the formalism established in Ref.~\cite{Picker:2023ybp}, where the MSBH population is parametrized by the mass $m$ of the produced black holes, and a fraction $\fMyr$ which describes the portion of the dark matter energy density which collapses into black holes in a Myr period. Then the `injection' rate is given by,
\begin{align}\label{eq:inj}
    \dndt(\mathbf{x}) = \frac{\fMyr}{m}~ \rho_{DM}(\mathbf{x})~.
\end{align}
This MSBH population should then reproduce the observed signal. A population of exploding black holes follows a `triangle' mass distribution, regardless of the initial mass function~\cite{carr_constraints_2016,Cai:2021fgm,mosbech_effects_2022,Picker:2023ybp} as in Ref.~\cite{Picker:2023ybp}. This is a result of the nonlinearity of the black hole mass loss equation~\cite{page1976ApJ...206....1P,mosbech_effects_2022}:
\begin{align}\label{eq:bhdm}
    \frac{dM}{dt} = -\frac{\hbar c^4}{G^2}\frac{\alpha_\mathrm{eff}(M)}{M^2}~.
\end{align}
In the above the effective factor $\alpha_\mathrm{eff}\sim10^{-3}-10^{-4}$ for small black holes accounts for the increase in permissible particle species black holes of smaller masses can emit. 

This triangle distribution---and the photon flux resulting  from this distribution---therefore is unique to exploding black holes. We show a possible realization of the differential flux from such an MSBH distribution in Fig.~\ref{fig:gcenter_spectrum}, where MSBHs of mass $m\sim10^{13}~g$ and an injection fraction per Myr of $\fMyr\sim2\times10^{-14}$ satisfy quite well the Fermi-LAT~\cite{fermilat2015ApJS..218...23A,Malyshev:2015hqa} observations of the Galactic Center excess, although we note that there are some analysis of the Fermi-LAT data which appear to support a higher-energy tail to the signal~\cite{Dinsmore:2021nip}. We also show Galactic Center observations by HESS~\cite{hess_collaboration_acceleration_2016}, VERITAS~\cite{Adams:2021kwm} and MAGIC~\cite{magic2020A&A...642A.190M} in the TeV range. These observations are already well described by the gamma rays produced from the acceleration of cosmic-ray protons to PeV energies in this region~\cite{HESS:2017tce,Adams:2021kwm}.

In addition to the spectrum, one must explain the morphology of the GeV excess, which continues to be the subject of debate in the literature. 
In a dark matter interpretation~\cite{Leane:2019xiy,Cholis:2021rpp}, the signal would more likely be consistent with a spherically symmetric source distribution around the Galactic Center, with a flux that falls off as $F_\gamma \propto r^{-(2.2-2.6)}$~\cite{Daylan:2014rsa}. Here the dark matter is expected to have a distribution $\rho(r) \propto r^{-\gamma}$, where $\gamma =1 $ in the absence of baryonic  contraction, and with the range $\gamma =0.6 - 1.3 $ in the presence of baryonic effects~\cite{Gnedin:2011uj}. However, some recent analyses~\cite{Bartels:2017vsx,Macias:2016nev,Calore:2021jvg} indicate that the signal may be better fitted to the Galactic bulge.

The rate of MSBH formation can have a nonlinear dependence on the dark matter density because $\fMyr$ in Eq.~\ref{eq:inj} can be a function of the dark matter density. We can write this generically as,
\begin{align}\label{eq:rdep}
    m(r) &= m_0 \left(\frac{\rhodm(r)}{\rhodm(r_0)}\right)^{p_{m}}\nonumber \\
    \fMyr(r) &= f_{\mathrm{Myr},0} \left(\frac{\rhodm(r)}{\rhodm(r_0)}\right)^{p_{f}}~,
\end{align}
where the index $0$ denotes a value at Earth's Galactic radius, and we are using a Navarro-Frenk-White (NFW) dark matter density profile~\cite{Navarro:1995iw}, with local density $\rho_0=0.01~\mathrm{M}_\odot~\mathrm{pc}^{-3}$ and scale radius $R_s=20~\mathrm{kpc}$. It is hard to imagine a scenario, however, where $m$ has position dependence---i.e., the dark structures collapse at different thresholds depending on the dark matter density---while also keeping $\fMyr$ constant. More likely, $\fMyr$ would also change with $m$ in such a scenario. 

On the other hand, it is not hard to envision scenarios where $m$ is constant but $\fMyr$ depends on the dark matter density. If the power $f_p\sim1$, we could reproduce the signal morphology as in Ref.~\cite{Daylan:2014rsa}, and some possible realizations of this are discussed below. The spectrum in Fig.~\ref{fig:gcenter_spectrum} assumes that there is such a radial dependence on $\fMyr$. We note that the magnitude of the signal is significantly larger than the equivalent in Ref.~\cite{Picker:2023ybp} as a result of this radial dependence, implying tighter galactic center constraints than previously calculated.

There is also the matter of whether the GeV excess appears to be more point-like or diffuse~\cite{Lee:2015fea,Calore:2018sbp,Leane:2019xiy,Dinsmore:2021nip}, although evidence that the signal is point-like is disputed~\cite{Leane:2019xiy, leanePhysRevLett.125.121105,Cholis:2021rpp}. While the large population of exploding black holes near the Galactic Center would likely best fit a smooth signal, most of the mechanisms discussed in the later sections of this paper for reproducing the $\sim\rho^2$ morphology involve increasing the likelihood of finding an MSBH closer to the Galactic Center. As a result, there would always be a smaller population of exploding black holes closer to Earth along the Galactic Center line of sight (perhaps in areas of increased dark matter density), which might make the signal less smooth---this would be harder to account for in the annihilating dark matter scenario, where the $\sim \rho^2$ dependence is required for the signal to be produced at all.

Possible realizations of MSBH collapse scenarios exist in the literature---Refs.~\cite{flores_primordial_2021,chakraborty_formation_2022,lu_late-forming_2022,kawana_primordial_2022,Domenech:2023afs} describe scenarios in which dark matter might form into microstructures which could collapse at late times. These formation scenarios rely on a large number of essentially free parameters and theoretical uncertainties---despite this however, it probably does not require significant fine-tuning to source a population of dark structures with a sufficiently extended mass distribution such that some of the structures are collapsing to MSBH today. Importantly, there is also a significant amount of room in the allowed parameter-space which could potentially source the important $\rho^2$ radial dependence of the Galactic Center signal. To quantitatively explore these possibilities, we would presumably need detailed numerical simulations of the dark structures from early times all the way through to today, which is well beyond the scope of our current understanding.  However, we can still propose a number of plausible scenarios that could well explain the radial dependence:

{\it 1. Microstructure distribution.} Dark matter microstructures may form from long range Yukawa forces in the dark sector, as in Ref.~\cite{flores_primordial_2021,Domenech:2023afs}. In this scenario, it may be the case that more structures are formed in areas with greater dark matter density, such as in the overdense halos which eventually become galaxies. Then one would naturally have larger $\fMyr$ near the Galactic Center, while the microstructures farther out in the galaxy have either yet to collapse or have already collapsed.

{\it 2. Microstructure mass function.} In addition, the microstructure population can have a nontrivial mass function. For instance, the higher dark matter densities can cause more fragmentation, leading to smaller microstructures. Or, depending on the particular realization, the opposite could be the case---namely, larger dark matter densities could source larger microstructures. In either case, the epoch of collapse will depend on distance from the Galactic Center, so the resulting  population of microstructures could have the masses of present-day collapse to black holes consistent with the required morphology. 

{\it 3. Microstructure density.} The halos forming in the denser dark matter regions could be denser than halos formed in less dense region (even if the mass is the same for both). In this case, if the halos cool volumetrically as in Ref.~\cite{flores_primordial_2021,lu_late-forming_2022,Domenech:2023afs}, the denser halos might collapse at earlier times than less dense halos, again setting up a radial dependence.

{\it 4. Microstructure clustering.} The same long-range Yukawa forces which create the microstructures might also provide an attractive force between the microstructures themselves~\cite{Flores:2021jas}. On the simplest level this could perhaps lead microstructures to cluster together in areas with greater dark matter density, where more microstructures have formed---i.e. in the Galactic Center. Furthermore, formation of microstructure binaries held together by Yukawa forces, similar to how PBHs might form early universe binaries could be a contributing factor~\cite{Sasaki:2016jop}. If the binary lifetime is long enough to delay the collapse until the present time, this could also introduce the required dependence on the dark matter density.


Finally, we must ensure that the MSBH abundance constraints of Ref.~\cite{Picker:2023ybp} are satisfied, particularly from the diffuse extragalactic emission, including now the effect of differing values for the radial distribution indices $p_{m,f}$. The diffuse constraints are calculated under the assumption of an isotropic dark matter background, with no reference to the morphology of distant galaxy halos, so it is not completely trivial to calculate their effect on the diffuse GeV signal, and each of our proposed mechanisms for the realization of this radial dependence actually have different consequences for the extragalactic diffuse constraints.

For instance, if we observe more MSBH explosions in the galactic center because the MSBH mass function itself has radial dependence, or if there is a density- but not mass-dependence (scenarios 2 and 3), the diffuse signal would be either smaller or the same as in the no-radial dependence case, depending on whether the galactic center MSBHs are exploding before, or after, the more distant black holes. In the scenario where microstructures formed uniformly but migrated in some way towards the galactic center (scenario 3), the diffuse constraints would not be affected in any way. However, if the radial dependence is due to an increased \textit{likelihood} of forming dark matter microstructures in more dense regions (scenario 1), the situation is more complicated.

In this scenario we can estimate an effective extragalactic, isotropic injection fraction $\widetilde{f}_\mathrm{Myr}$ given that the the local injection fraction now has radial dependence within each galaxy, as in Eq.~\ref{eq:rdep}. This effective injection fraction can then be used to read off the limits from Ref.~\cite{Picker:2023ybp}. We roughly estimate this by averaging $\fMyr(r)$ over the volume of a galactic halo, assuming that the majority of the dark matter is in galactic halos (this is generally a conservative assumption, since dark matter not in halos is likely to be less dense and so have smaller injection fraction in this scenario). Then the effective global injection fraction is,
\begin{align}
    \widetilde{f}_\mathrm{Myr} &\sim\frac{4\pi}{(4/3)\pi R_\mathrm{vir}^3}\int_0^{R_\mathrm{vir}} \fMyr(r) r^2 \mathrm{d}r~.
\end{align}
For simplicity we take the Milky Way NFW parameters as representative of an average galaxy and use the virial radius $R_\mathrm{vir}\sim 10 R_s$ as the maximum halo extent. For the relevant case $p_f=1$, we then find,
\begin{align}
    \widetilde{f}_\mathrm{Myr} \sim 0.04 f_{\mathrm{Myr},0}~.
\end{align}
We can understand this result simply by noting that there is more dark matter outside Earth's galactic radius than interior to it, although though it is less dense. As a result the global injection fraction is smaller than the value at Earth's galactic radius. For the value $f_{\mathrm{Myr},0}\sim2\times 10^{-14}$ which reproduces the GeV excess, we then find that we entirely avoid the diffuse constraints from our previous work~\cite{Picker:2023ybp} for any length of injection period shorter than the age of the universe.

In summary, we showed that evaporating black holes which form from collapsing dark matter microstructures can produce a gamma-ray signal with the correct spectrum to explain the GeV excess and additionally outlined the generic requirements for such a scenario to adequately capture the signal. We have also outlined the criteria for MSBH formation to explain the signal morphology. Explicit scenarios inspired by recent literature~\cite{flores_primordial_2021,lu_late-forming_2022,Domenech:2023afs} will be presented elsewhere, and more sophisticated statistical analyses should follow when the stronger theoretical predictions exist. 


This paper was partially written on Tongva, Chumash, Eora, and Massachussett lands. We would like to thank Markus Mosbech for providing BlackHawk data from our previous work~\cite{mosbech_effects_2022}.  We also thank K.~Kawana and P.~Lu for helpful discussions. 
This work was supported by the U.S. Department of Energy (DOE) Grant No. DE-SC0009937. The work of A.K. was also supported by World Premier International Research Center Initiative (WPI), MEXT, Japan, and by Japan Society for the Promotion of Science (JSPS) KAKENHI Grant No. JP20H05853. 

This work made use of N\textsc{um}P\textsc{y}~\cite{numpy2020Natur.585..357H}, S\textsc{ci}P\textsc{y}~\cite{scipy2020NatMe..17..261V}, and M\textsc{atplotlib}~\cite{mpl4160265}.
\bibliographystyle{bibi}

\bibliography{late_pbh.bib}
\end{document}